\documentclass[12pt,a4]{iopart}

\usepackage{graphicx}
\usepackage{url}
\usepackage[dvips]{hyperref}

\newif\ifpdf
\ifx\pdfoutput\undefined
\pdffalse 
\else
\pdfoutput=1 
\pdftrue
\fi

\begin{document}

\title[Characterization of coherent backgrounds in continuously acquired data]
{A method for characterization of coherent backgrounds in real time and
its application in gravitational wave data analysis}

\author{E.\,J.~Daw\footnote{To whom all correspondence should be addressed}}
\address{University of Sheffield, Department of Physics and Astronomy, 
Hicks Building, Hounsfield Road, Sheffield S3 7RH, UK}
\ead{e.daw@shef.ac.uk}

\author{M.\,R.~Hewitson}
\address{Max-Planck-Institut f\"ur Gravitationsphysik (Albert-Einstein-Institut),
Callinstrasse 38, D-30167, Hannover, Germany}

\begin{abstract}
Many experiments, and in particular gravitational wave detectors, produce
continuous streams of data whose frequency representations contain discrete,
relatively narrowband coherent features at high amplitude.
We discuss the application of digital Fourier transforms (DFTs) to characterization
of these features, hereafter frequently referred to as lines. Application of DFTs
to continuously produced time domain data are achieved through
an algorithm \cite{halberstein},
hereafter referred to as EFC\footnote{Evolving Fourier coefficient},
for efficient time-domain determination of the Fourier coefficients of a data set. 
We first define EFC and discuss
parameters relating to the algorithm that determine its properties and action
on the data. In gravitational wave interferometers, these lines are commonly
due to parasitic sources of coherent background interference coupling
into the instrument. Using GEO\,600 data, we next demonstrate that 
time domain subtraction of lines can proceed without detrimental effects either on features at frequencies separated from that of the subtracted line, or on features at the 
frequency of the line but having different stationarity properties.
\end{abstract}


\section{Introduction}
\label{sec:introduction}

The purpose of this paper is to describe and validate a method for characterizing and 
removing coherent backgrounds, hereafter referred to as lines,
in continuously acquired data. There has been much work
in this area, including several studies of techniques
intended specifically for application to data from gravitational wave detectors. 
See for example references \cite{mohanty,finn,sintes,allen,motin}. These 
existing methods make very limited use of Fourier coefficients, because
the usual methods for obtaining these coefficients through the fast Fourier
transform (FFT) algorithm \cite{tukey}
run into practical difficulties where it is required that
the Fourier coefficients be equally accurate estimators of line parameters for
all data samples, and that the estimate of line parameters be sufficiently fast
to keep pace with the data sampling rate.

In this paper, we describe the application of an
alternative to the fast Fourier transform, referred to throughout the paper
as EFC, for obtaining the Fourier coefficients
continuously updated at the data sampling rate. This algorithm was
described in 1966 by Halberstein \cite{halberstein},
and is the subject of U.S. and E.U. patents. 
See \cite{uspatent,eupatent}, for example. The authors exploit this algorithm
for the monitoring of coherent backgrounds in gravitational wave detector data.
We demonstrate that EFC is significantly faster than real time running on
contemporary computers, and equally accurate for all data samples. We continue by providing a proof of principle of EFC using data from a ground-based gravitational wave detector, and by studying the response of the algorithm to injected signals.

A key application of a robust real time capable line removal tool is
the generation of triggers for networks of interferometric gravitational
wave detectors in real time for the purposes of providing data for 
follow up of triggers with optical or radio observations.
Applications of EFC in this area are more widespread than the removal
of lines. A continuously maintained monitor of magnitude and phase of 
coherent signals in the data has many applications in gravitational wave
interferometery, including calibration of the detectors, monitoring 
interferometer subsystems such as high Q suspensions, and studies
of bicoherence between lines and features at other frequencies in the 
same data. 

\section{Line parameterization with Fourier transforms}

By a line we mean an oscillating feature in the data whose
amplitude $A_i$ at the time, $t_i$, of the $i^{\mathrm{th}}$ sample
can be written

\begin{equation}
A_n=A_0\cos (2\pi f t_n + \phi),
\label{eqn:spectralfeature}
\end{equation}

where $A_0$, $f$, and $\phi$ are the underlying amplitude, frequency, and phase
of the spectral feature, which are functions of time $t$. The algorithm we will
describe requires that these parameters do not vary significantly on timescales
less than some cutoff $\tau$. We will also confine the discussion
to lines having $f(t)$ combined to a well defined domain for all $t$, and to 
data sampled regularly, meaning that the sampling time, $t_s=t_{n+1}-t_n$, is 
independent of $n$.

If the frequency $f$ is known precisely, it only remains to evaluate the 
amplitude $A$ and phase $\phi$ at each time $t_n$. A natural starting point
would be the $k^{\mathrm{th}}$ coefficient of the digital Fourier transform
\cite{oppenheim} of $N$ successive data samples $x_j$ centered on the 
$n^{\mathrm{th}}$ data sample.

\begin{equation}
\mathcal{F}_k(x_0,~x_1,\cdots,x_{N-1})=
\sum_{j=0}^{N-1} x_j e^{\frac{-2\pi i jk}{N}},
\label{eqn:fourier}
\end{equation}

where $k\simeq Nt_sf$ and $Nt_s\le \tau$. In the case where $f$ is not
precisely known, or corresponds to non-integer $k$, the values
of several Fourier coefficients would be needed.
Fourier coefficients can be extracted from the DFT of the data, which is 
frequently calculated using a fast Fourier transform (FFT) algorithm
\cite{oppenheim,numrec}. The number of floating point operations
to calculate the DFT of $N$ successive data points is of order
$N\log_2N$.

This method is beset with difficulties in the
case where one requires a continuous, sample-by-sample 
estimator of $A$, $f$ and $\phi$. Because fast Fourier transform
algorithms require a block of time series data, it is not obvious how to proceed from
an estimate of $A_n$ and $\phi_n$ at the $n^{\mathrm{th}}$ data sample to an
estimate of the $A_{n+1}$ and $\phi_{n+1}$.
It is too computationally intensive to perform an FFT
for each time sample - $N$ samples would
require order of $N^2\log_2N$ operations. The alternative usually
considered is performing a new FFT every $m$ samples,
where $m<N$. The problem in this case
is the occurrence of discontinuities in the estimator at each new transform.

\section{The evolving Fourier coefficient (EFC) algorithm}
\label{sec:efc}

In the case under discussion, only a restricted number
of Fourier coefficients are of interest, suggesting an alternative
approach. One starts by 
generating Fourier coefficients of $N$ data points using
Equation \ref{eqn:fourier}. 
It is shown in \ref{sec:appendix} that the $k^{\mathrm{th}}$
Fourier coefficient, $\mathcal{F}_k(x_1,~x_2,\cdots,x_{N})$, of
time samples $x_1,~x_2,\cdots,x_N$ is related to 
$\mathcal{F}_k(x_0,~x_1,\cdots,x_{N-1})$ by

\begin{equation}
\mathcal{F}_k
(x_1,~x_2,\cdots,x_{N})=
e^{\frac{+2\pi i k}{N}}
\left(
\mathcal{F}_k(x_0,~x_1,\cdots,x_{N-1})
+(x_N-x_0)
\right).
\label{eqn:mft}
\end{equation}

\begin{figure}[htbp]
\begin{center}
\includegraphics[width=0.8\textwidth]{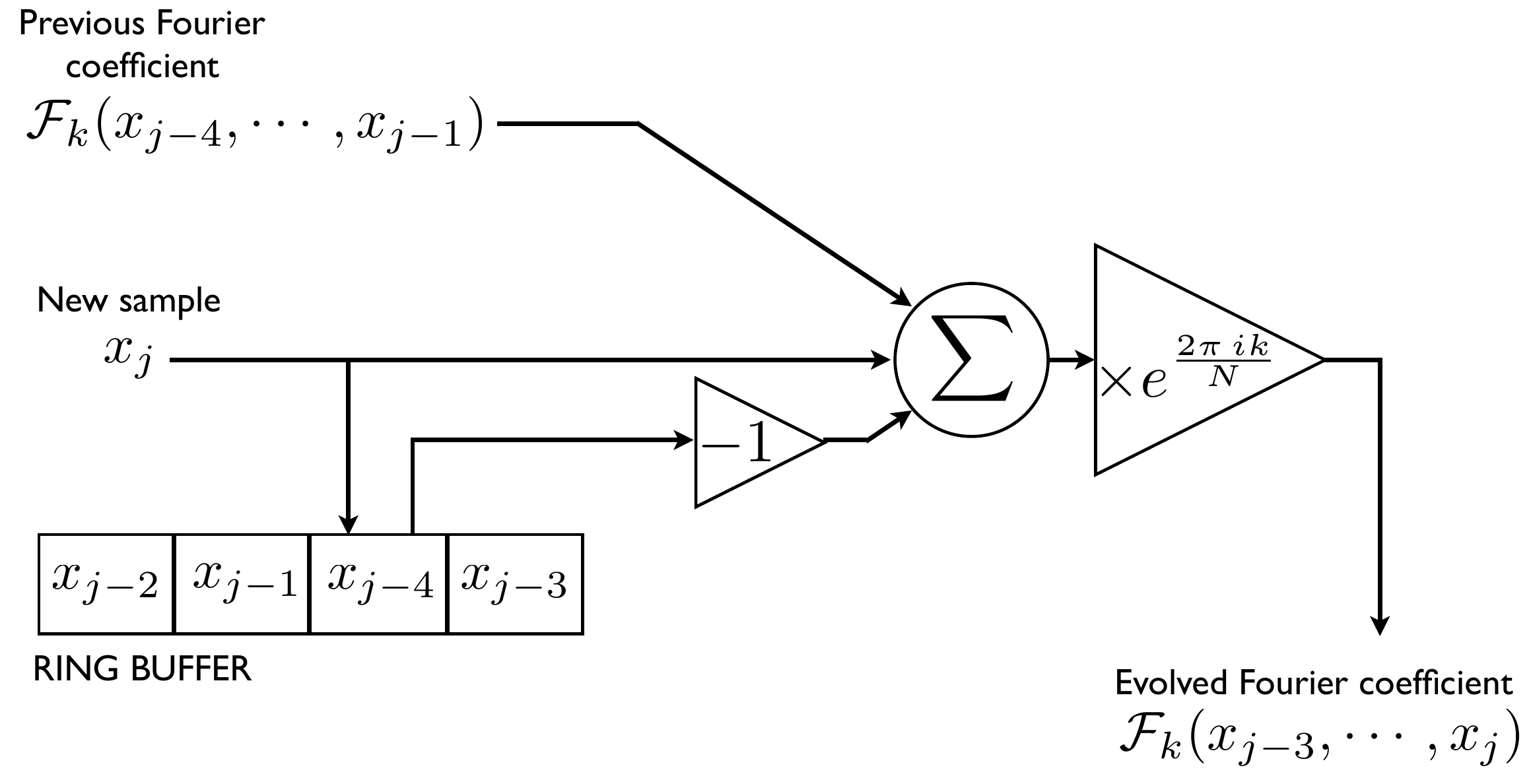}
\caption{\label{fig:efcblockdiagram} A block diagram illustration of the 
operation of the EFC algorithm for the simple case where each FFT is 
over four data samples. In the example iteration
illustrated here, the sum of the new data sample
and minus a data sample taken four sampling times previously are added
to the previous Fourier coefficient, and the result of this sum is multiplied
by $\exp(i\delta)$, where $\delta=2\pi k/N$ is the phase difference between successive
time samples at the frequency of the Fourier component. The newest data
sample then replaces the fifth most recent data sample in the ring buffer.}
\end{center}
\end{figure}

Using this result, the computation of the Fourier coefficient of
the overlapping time series is achieved without the need for an
additional Fourier transform. The process can be repeated to
obtain the time evolution of the Fourier coefficient for all 
subsequent time samples. Figure \ref{fig:efcblockdiagram}
illustrates the algorithm for the
case where the sum is over four successive data samples.

We recall some properties of discrete Fourier transforms whose
choice will be important in practical application of this technique
to line monitoring and removal. First, the Fourier coefficient determines
the line parameters over a data set of duration $\tau=Nt_s$. This timescale
should be shorter than the typical timescale for variations in the line
parameters $A$, $f$ and $\phi$, but longer than the typical timescale
for features in the data that you would not wish to influence the line
parameterization. The duration $\tau$ also determines the resolution bandwidth
$B=1/\tau$, the bandwidth in frequency space sampled by a single Fourier 
coefficient. If the lines under study
have a width or range of wander $\Delta f$, then the number of Fourier
coefficients to be calculated is $\tau\Delta f = Nt_s\Delta f$ per new time
sample. 

EFC requires an initial value for each Fourier coefficient
of the data. We note, however, that if we use zeros for the values of $x_0$ subtracted
in Equation \ref{eqn:mft} for the first $N$ timesamples, the output of the algorithm
exactly equals the Fourier component after $N$ iterations. Hence there is no need for
initial values of the Fourier components, since the iteration algorithm generates them.
Equation \ref{eqn:mft} implies that the algorithm must evaluate the difference between the most recent data value $x_N$ and the data value $x_0$ acquired $N+1$ sampling times
previously. After startup of the algorithm, the latter sample is unavailable until $N$ sampling times
have elapsed, hence there is an unavoidable start-up period of $Nt_s$ to initialize the 
filter during which the algorithm output is not useful for line characterization.
The response of the algorithm to transients is discussed in Section \ref{sec:signals}.
A natural and fast way to implement access to the $N+1$ most recent data samples
is to store $N$ previously acquired samples in
a ring buffer, where at each iteration the oldest stored sample
$x_0$ is read, applied in the algorithm, and then overwritten with the most recent sample
$x_N$. A counter is then incremented, modulo $N$, so that the next element in the ring
buffer, containing the input data value to be used as $x_0$ in the next iteration of
Equation \ref{eqn:mft}, is referenced.

\begin{figure}[htbp]
\begin{center}
\includegraphics[width=1.0\textwidth]{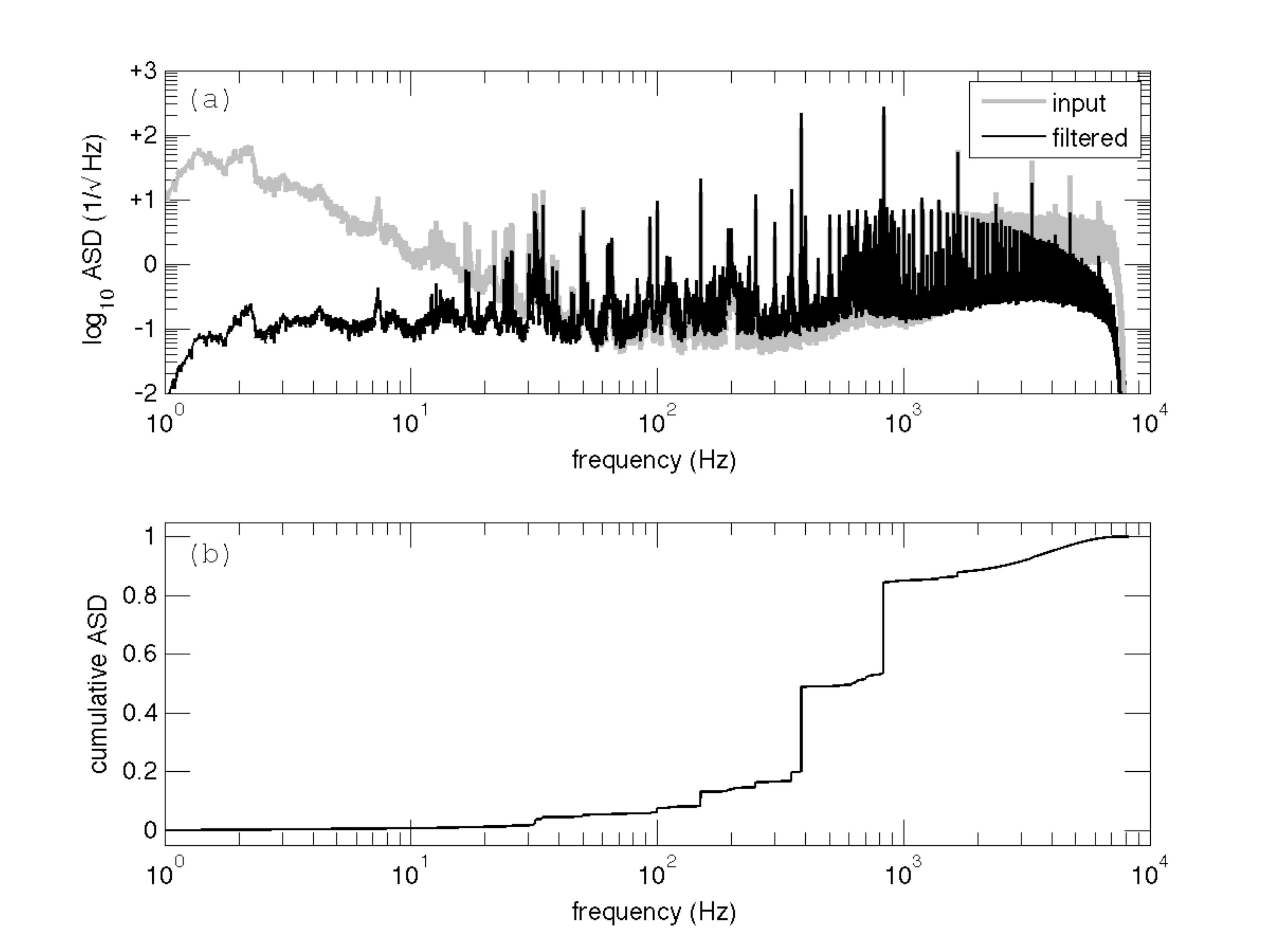}
\caption{\label{fig:untreatedasd} Amplitude spectral density (ASD) of noise from
the GEO\,600 michelson photodiode channel \texttt{G1:LSC\_MIC\_VIS}. The upper plot shows amplitude per $\surd$Hz
as a function of frequency, where each bin has a bandwidth of 7.8\,mHz.
The dashed and solid curves are 
of data before and after the whitening prefilter discussed in Section \ref{sec:thedata}.
The lower plot is the cumulative ASD of the whitened data,
obtained by adding in quadrature successive bins
from the bold curve in the upper plot and normalizing to the maximum value.
Lines contributing significantly 
to the time domain amplitude of the signal are shown as large, discrete steps
in the cumulative ASD.}
\end{center}
\end{figure}

\section{Tests on GEO\,600 data}
\label{sec:geo}

\subsection{Characteristics of the test data}
\label{sec:thedata}

The immediate application for which the algorithm was conceived is the 
removal of coherent noise backgrounds from data produced by gravitational
wave interferometers, such as the GEO \cite{Willke:2002bs}, LIGO \cite{ligo},
VIRGO \cite{virgo} and TAMA \cite{tama}
instruments. For a general review of the science of direct gravitational wave detection the reader is referred to \cite{gwreview}.

The tests of EFC described in this paper were carried
out on data acquired from the GEO\,600
gravitational wave interferometer through the LSC-Virgo scientific collabration.
The GEO600 detector has been described in detail
elsewhere \cite{Willke:2002bs}. For the purposes of testing this algorithm, 
an arbitrary 1000 seconds of data from the \texttt{G1:LSC\_MIC\_VIS} channel,
sampled at 16384~Hz, and acquired when the instrument was in maintenance
mode, was used. This signal represents a measurement of the
DC light power reflected from the Michelson at the input port, commonly
referred to as the common-mode visibility. 

Figure \ref{fig:untreatedasd} shows the amplitude spectral density (ASD) of the
data before filtering. Plot (a) shows power spectral density over 7.8\,mHz
bins from 1~Hz to 8192~Hz, the Nyquist frequency of the data acquisition system.
The dotted curve on this plot shows the ASD of the raw data. Broadband noise below
50\,Hz and above 1\,kHz dominates the cumulative amplitude spectral density. 
These components are suppressed using an IIR whitening filter \cite{oppenheim}
consisting of two zeros at 0~Hz, two real poles at 50~Hz and a single real pole at 1\,kHz. 
The solid curve is an ASD of the data after this preprocessing step. 

Plot (b) from figure \ref{fig:untreatedasd} shows a cumulative ASD, obtained
by squaring the ASD of the prefiltered data, integrating over frequency, taking 
the square root of the result, and dividing by the contents of the highest frequency
bin. Line features in the data making the largest contributions to the overall signal 
amplitude appear as steps in the cumulative ASD. Table \ref{tab:lines} column 1 shows the
frequencies of some of these lines.

\begin{table}
\begin{center}
\caption{\label{tab:lines}Performance of the subtraction algorithm for the fifteen
lines treated in the test. At each frequency (column 1), a number of bins (column 3)
corresponding to a bandwidth (column 2) about the line centre were treated with 
the subtraction algorithm. After subtraction, the percentage shown in column 4 
of the line amplitude before subtraction remained. Line amplitudes were determined
using the frequency bin having the maximum ASD before subtraction.}
\vspace{3mm}
\begin{tabular*}{0.85\textwidth}{p{30mm}p{30mm}p{30mm}p{30mm} }
Frequency (Hz) & Bandwidth (Hz) & Number of bins & Remnant (\%) \\
\mr
150 & 1 & 8 &  7.1 \\
250 & 1 & 8 & 8.5 \\
350 & 1 & 8 & 8.8 \\
384 & 2 & 16 & 1.3 \\
550 & 1 & 8 & 14 \\
614 & 5 & 40 & 1.3 \\
650 & 1 & 8 & 16.3 \\
709 & 5 & 40 & 1.2 \\
803 & 2 & 16 & 9.4 \\
828 & 5 & 40 & 0.44 \\
1391 & 2 & 16 & 3.8 \\
1654 & 5 & 40 & 0.79 \\
3308 & 2 & 16 & 24 \\
\mr
\end{tabular*}
\end{center}
\end{table}

\subsection{Line removal}
\label{sec:lineremoval}

The practical application of equation \ref{eqn:spectralfeature} to remove lines in the data
is to estimate the amplitude and phase of oscillatory components of the preprocessed
data at the frequencies of identified lines, to synthesize an oscillation having
this amplitude and phase, and to subtract this oscillation from the data.
The first step on receiving a new sample of data $x_k$ is to use equation
\ref{eqn:spectralfeature} to generate an updated Fourier component $y_k$. The real part
of this Fourier component is proportional to the wave of frequency $k/Nt_s$ contributing to
the data. Note, however that the phase of $y_k$ represents the phase at the
beginning of the timeseries,
whereas for subtraction we require the phase half way through the timeseries
\footnote{In practice the time series has an even number of samples, so we take the
subtraction to occur at sample k-N/2, which is the leftmost of the two samples in the 
centre of the time series}. Therefore $y_k$ is modified by a phase factor
representing the phase shift over $N/2-1$ samples, and then the real part of the result
is taken. The correction signal is denoted $s_k$, and is given by

\begin{equation}
s_k = \frac{2}{N}\Re (y_ke^{i\pi k\left(1-\frac{2}{N}\right)}),
\label{eqn:synthesized}
\end{equation}

where $N$ and $y_k$ have the same meaning as in Section \ref{sec:efc} and $\Re$ denotes
the real part. One final important detail is the time lag between the data and the correction. 
$y_k$ is the Fourier component of input data $x_{k-N+1}\cdots x_k$, therefore it best
represents the amplitude and phase of the corresponding frequency component at data
sample $k-N/2$. Therefore the actual subtraction of the synthesized wave is performed
on a timesample $N/2$ sample times lagging behind the $k^{\mathrm{th}}$ input sample. In
this sense the line subtraction filter is acausal; it requires data both before and after the 
point of interest to determine the correction at that point. For post-processing removal of lines,
for example as part of an off-line analysis chain, this is not a problem. The 
line subtracted data set $\{r_k\}$ is given by 

\begin{equation}
r_{k-N/2}=x_{k-N/2}-s_k
\label{eqn:subtraction}
\end{equation}

for all $k$. Note that the above equation subtracts a synthesized wave representing only one
Fourier component from the data set. The method can be applied to multiple components,
and to lines spanning more than one bin in Fourier space. The bandwidth of adjacent 
frequency bins is the reciprocal of the time duration of the equivalent Fourier transform.

\subsection{Computational overhead of the processing algorithm}
\label{sec:speed}

A practical line removal algorithm must be able to process the data at a rate at
least as high as the production rate for the data set. Fast data channels from existing
gravitational wave detectors sample at $t_s=1/16384~\mathrm{s}$.  The algorithm
was implemented in the programming language C,
and benchmarked on a MacBook Pro having a clock speed
of 2.33~GHz  and an Intel Core 2 Duo chipset, using the ANSI C time library and the
computer's on-board clock.
The time per incoming data sample per Fourier component analyzed
was 1/1700 of $t_s$, so the code can process 1700 Fourier components in parallel
on a single CPU  and match the incoming data rate. Because each Fourier 
component is treated separately, the algorithm could be run on a cluster in which case
the number of frequencies treated is limited only by the number of cluster nodes.
Tests on an intel based PC running linux yielded 1730 Fourier components in 
$t_s$.

\begin{figure}[htbp]
\begin{center}
\includegraphics[width=0.8\textwidth]{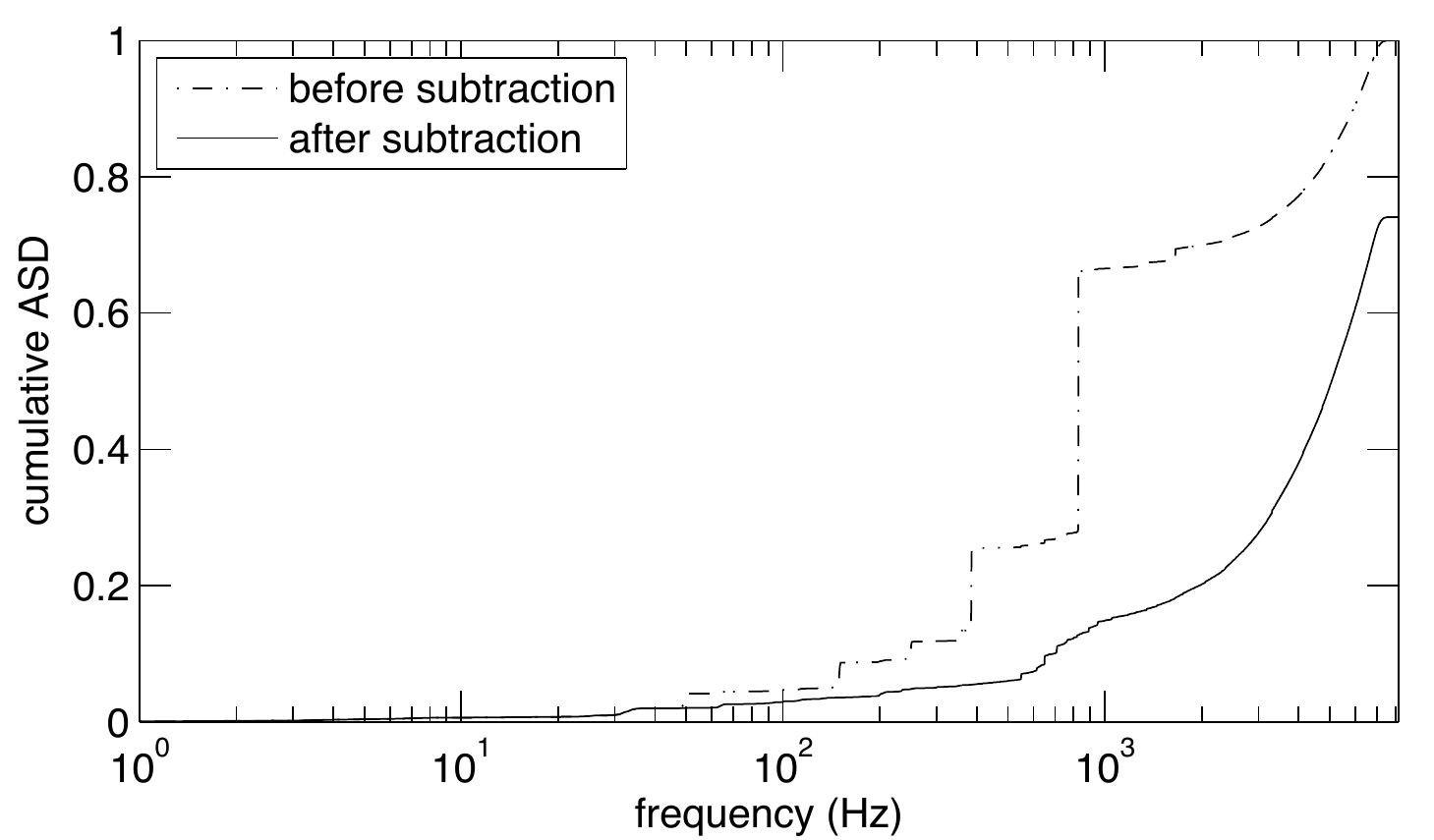}
\caption{\label{fig:asdbefaft} Cumulative amplitude spectral density (ASD) of noise from
the GEO\,600 michelson photodiode channel \texttt{G1:LSC\_MIC\_VIS} 
before and after subtraction of the lines listed
in Table \ref{tab:lines}. Both curves are scaled to the maximum of the cumulative
amplitude spectral density of the unsubtracted data. At the Nyquist frequency, the
amplitude of the line subtracted data is 0.54 that of the unsubtracted data.}
\end{center}
\end{figure}

\begin{figure}[htbp]
\begin{center}
\includegraphics[width=0.8\textwidth]{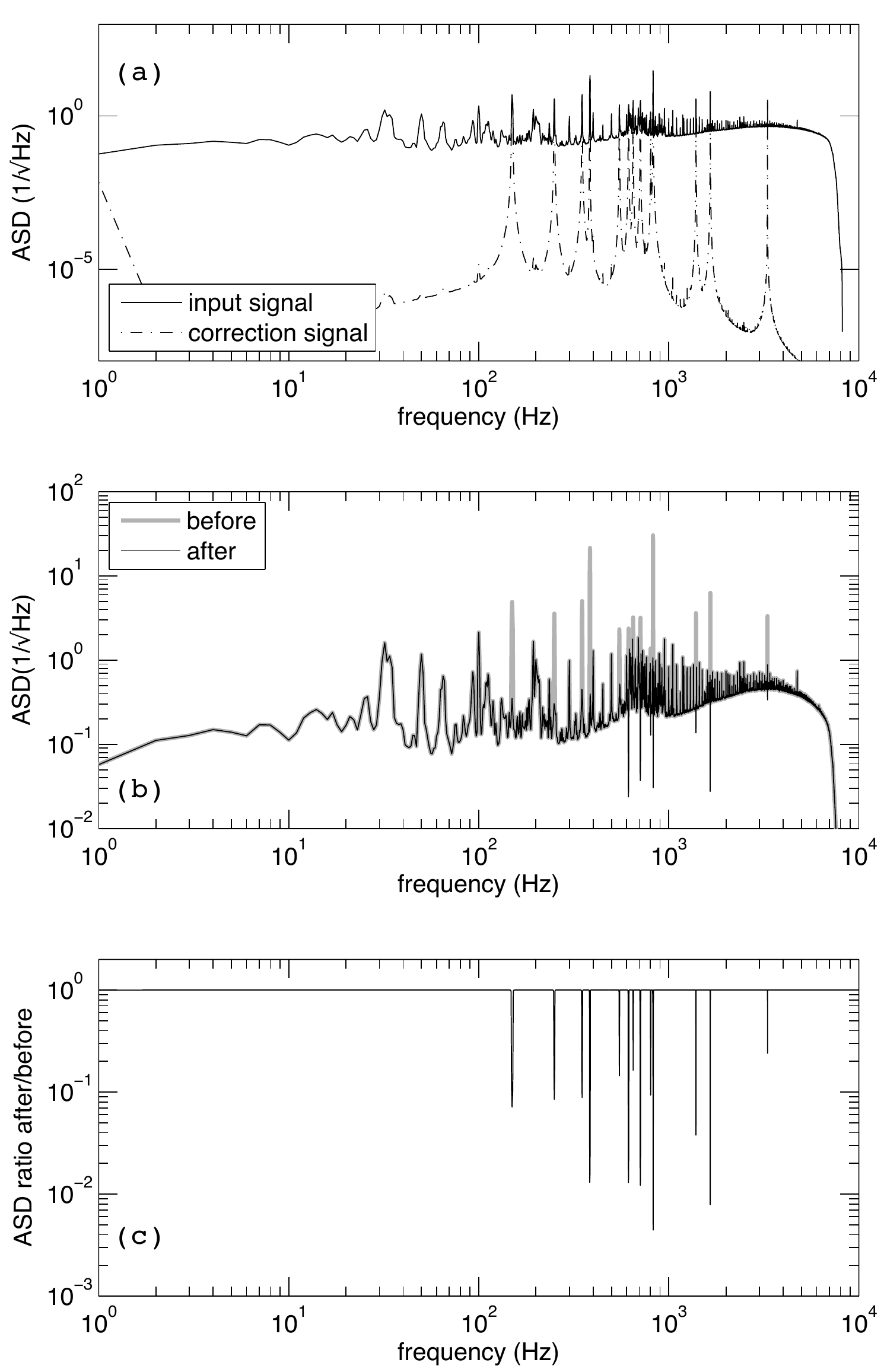}
\caption{\label{fig:asdratio} ASD of noise before
line subtraction with the amplitude spectral density of the subtracted
waveform overlaid - plot (a). ASD of the noise before (black) and after (grey)
line subtraction - plot (b). Ratio of the ASD after to before line subtraction versus frequency - plot (c). The efficiency of line removal, defined as one minus the ratio of the ASD after to before at the line peak, is better than
76\% for all the lines.}
\end{center}
\end{figure}

\begin{figure}[htbp]
\begin{center}
\includegraphics[width=0.9\textwidth]{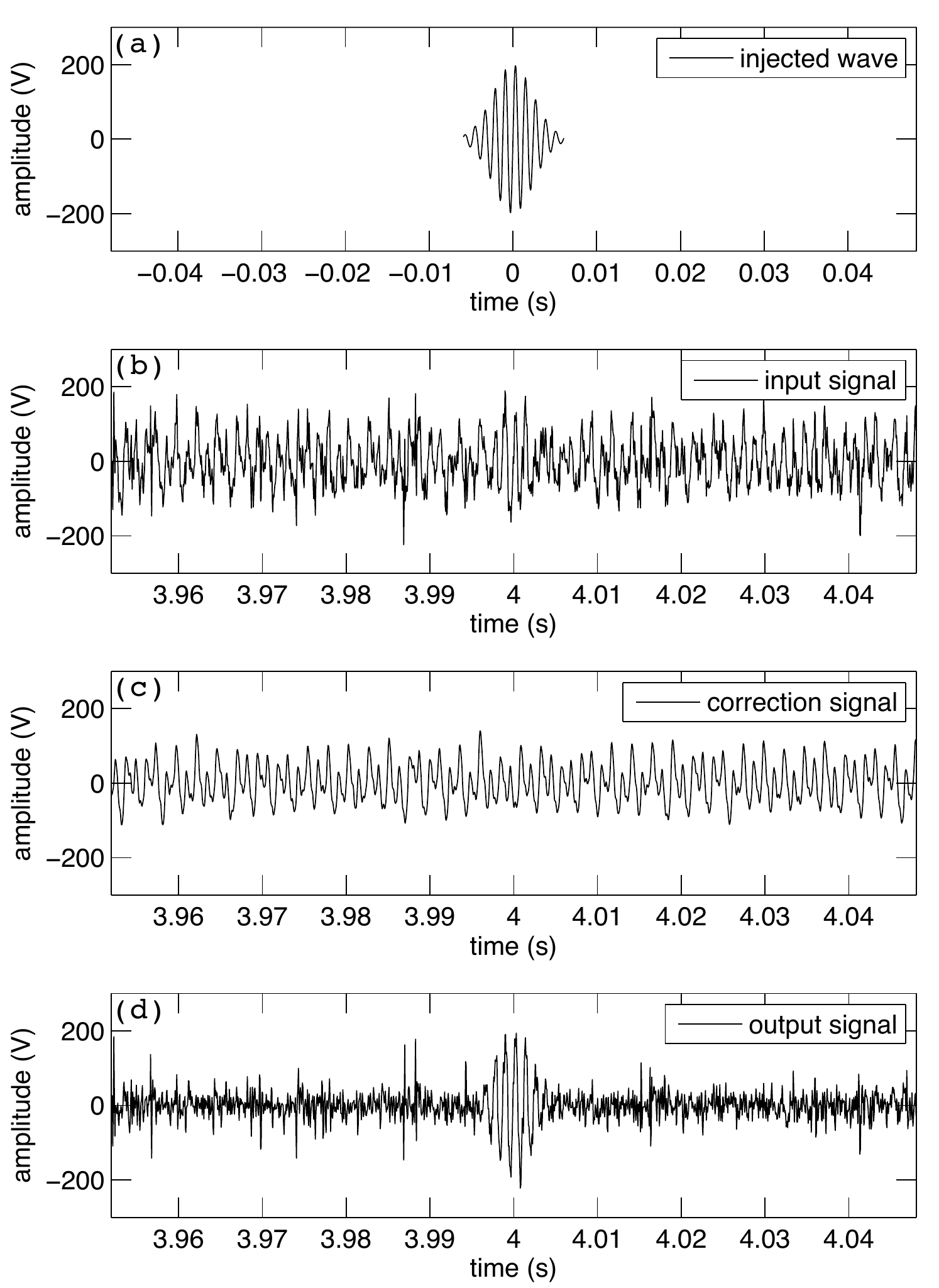}
\caption{\label{fig:timesig} Time series 50\,ms of data before (b),
and after (d) the subtraction of the correction signal (c)
generated using the MFT algorithm. So (b) is the sample-by-sample
sum of (c) and (d). A sine-gaussian having the same sine wave frequency
as the subtracted line (828\,Hz), shown in plot (a), was added to the data in plot (b)
to show that even bursts at line frequencies are unaffected by line 
subtraction as long as their duration is much less than the response 
time of the filter.}
\end{center}
\end{figure}

\subsection{Performance on the test data}
\label{sec:performance}

Performance on the test data is summarized in Figures \ref{fig:asdbefaft}, \ref{fig:asdratio},
and \ref{fig:timesig}. Figure \ref{fig:asdbefaft} shows the cumulative ASD
before and after subtraction on the same scale. The r.m.s. amplitude of the prewhitened
data is reduced to 0.54 of its unprocessed value after line removal. Figure \ref{fig:asdratio} shows the
ASD of the input data with the ASD of the correction superimposed; as expected the 
correction signal is dominated by sharp peaks at the frequencies of each of the corrected
lines. Plot (b) of this figure shows the ASD of the data before and after line subtraction
overlaid. Plot (c) shows the ratio of the bin-by-bin ASDs of the data
after to before line removal; each line is removed to a maximum of 24\% of its pre-treatment
value, with the largest lines (having the highest SNR in the power spectrum) attenuated to 0.6\% of their amplitude in the input data. The 4\textsuperscript{th}
column of Table~\ref{tab:lines} shows the percentage of the pre-subtraction ASD remaining at the Fourier component dominating the line after subtraction. These numbers correspond to the depth of the dips at the line frequencies in Figure \ref{fig:asdbefaft}, plot (c). 

\subsection{The effect of a noise background}
\label{sec:noise}

Noise in the subtracted signal enters through the noise background in the frequency
bin being monitored. This noise background is considered here as broadband Johnson
noise having a flat amplitude spectral density across the bin. As discussed in Section
\ref{sec:efc}, the bandwidth of a single bin is given by $B=1/\tau$ where $\tau$ is the 
duration of the timeseries used to estimate each Fourier coefficient. If the amplitude of
the signal is $A_S$ and the noise has an amplitude spectral density
$\lambda_N~/\surd{\mathrm{Hz}}$,
then the signal to noise ratio in power, $\mathrm{snr}$ is given by

\begin{equation}
\mathrm{snr}=\frac{A_S^2\tau}{\lambda_N^2} 
\label{eqn:snr}
\end{equation}

If subtracting the line with a given $\tau$ leads to addition of noise, the only recourse
is to increase $\tau$, and therefore decrease the bandwidth sampled per Fourier
coefficient. In the case where the line covers several bins in frequency space, the number
of Fourier coefficients subtracted will rise as you increase $\tau$, resulting in an increase
in CPU usage proportional to the increase in $\tau$. 
In addition, larger $\tau$ increases the response time of the filter. As discussed in Section
\ref{sec:signals}, this will make the subtraction insensitive to a greater range of 
time durations of bursts, which may be desirable from the point of view of signal
searches. On the other hand, $\tau$ should not be so large that it exceeds the timescales
typical of fluctuations in the amplitude or phase of a Fourier component of the line. The authors found
that $\tau$ of the order of 8-32 seconds satisfied these requirements for the lines detailed
in Table \ref{tab:lines}.
Relating signal to noise ratio to efficiency of the line remover, lower signal to noise ratio
leads to less efficiency in line removal. If the frequency of the line is stable and its parameters
are slowly varying, one can still increase efficiency by increasing $\tau$. Otherwise the line
subtraction is unavoidably inefficient. However, where the main aim of this tool is to 
suppress lines that make significant contributions to the cumulative ASD, and therefore
to the time domain amplitude of the data stream, we observe that all
lines corresponding to discernable steps in the cumulative ASD have been efficiently
removed after subtraction.

\begin{figure}[htbp]
\begin{center}
\includegraphics[width=0.6\textwidth]{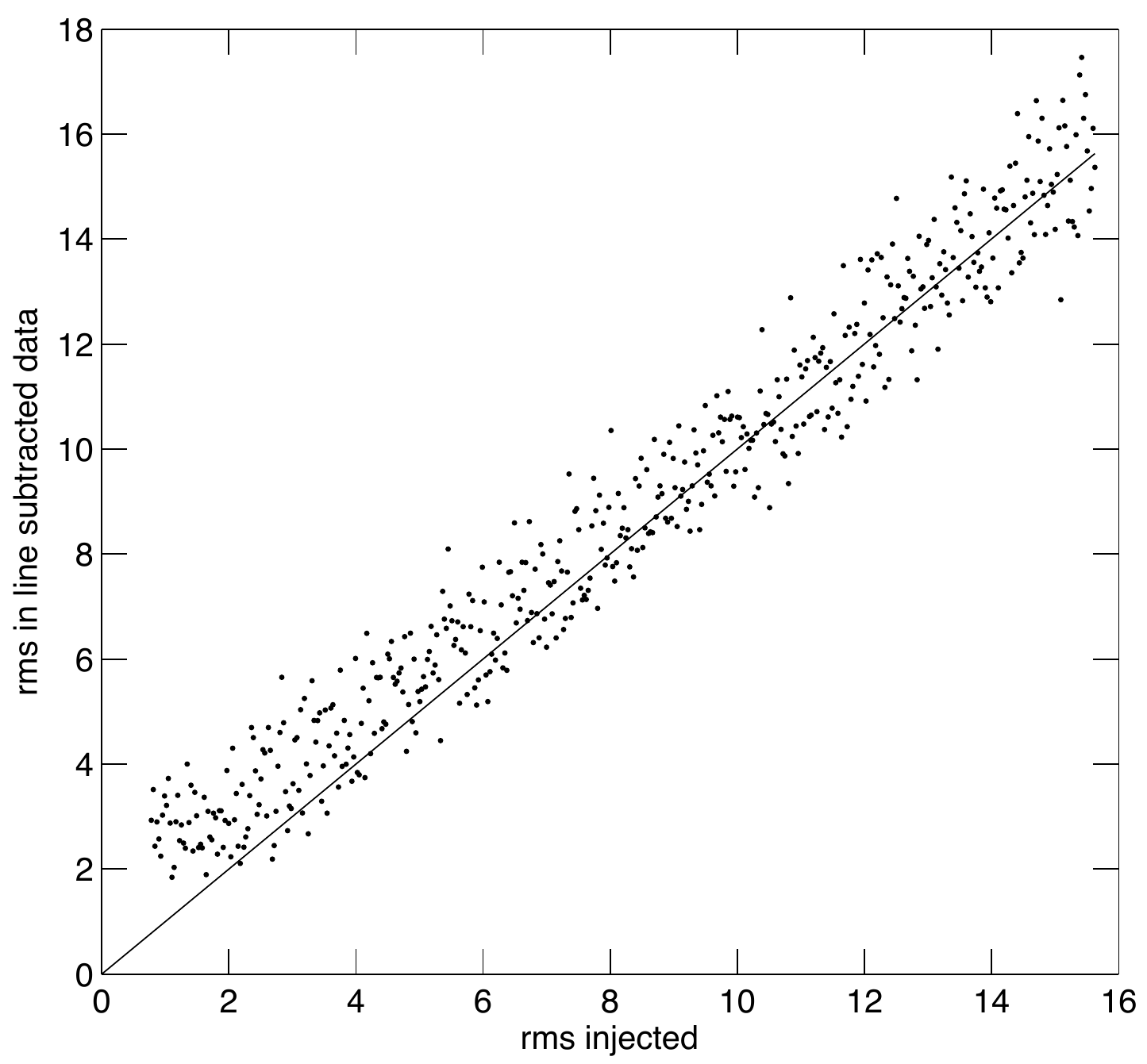}
\caption{\label{fig:multiinject} Amplitude of sine gaussian signals
after they have been injected at random positions in a 256 second
segment of data, and after the line remover has been run, versus
the amplitude of the signal before injection. Both amplitudes are
expresssed in units of the r.m.s. of the data before injections.
There were 500 injections with amplitudes of between 0.78 and
15.23 times the r.m.s. noise level. As a guide to the eye, a straight
line of slope one through the origin is also plotted.}
\end{center}
\end{figure}

\begin{figure}[htbp]
\begin{center}
\includegraphics[width=0.8\textwidth]{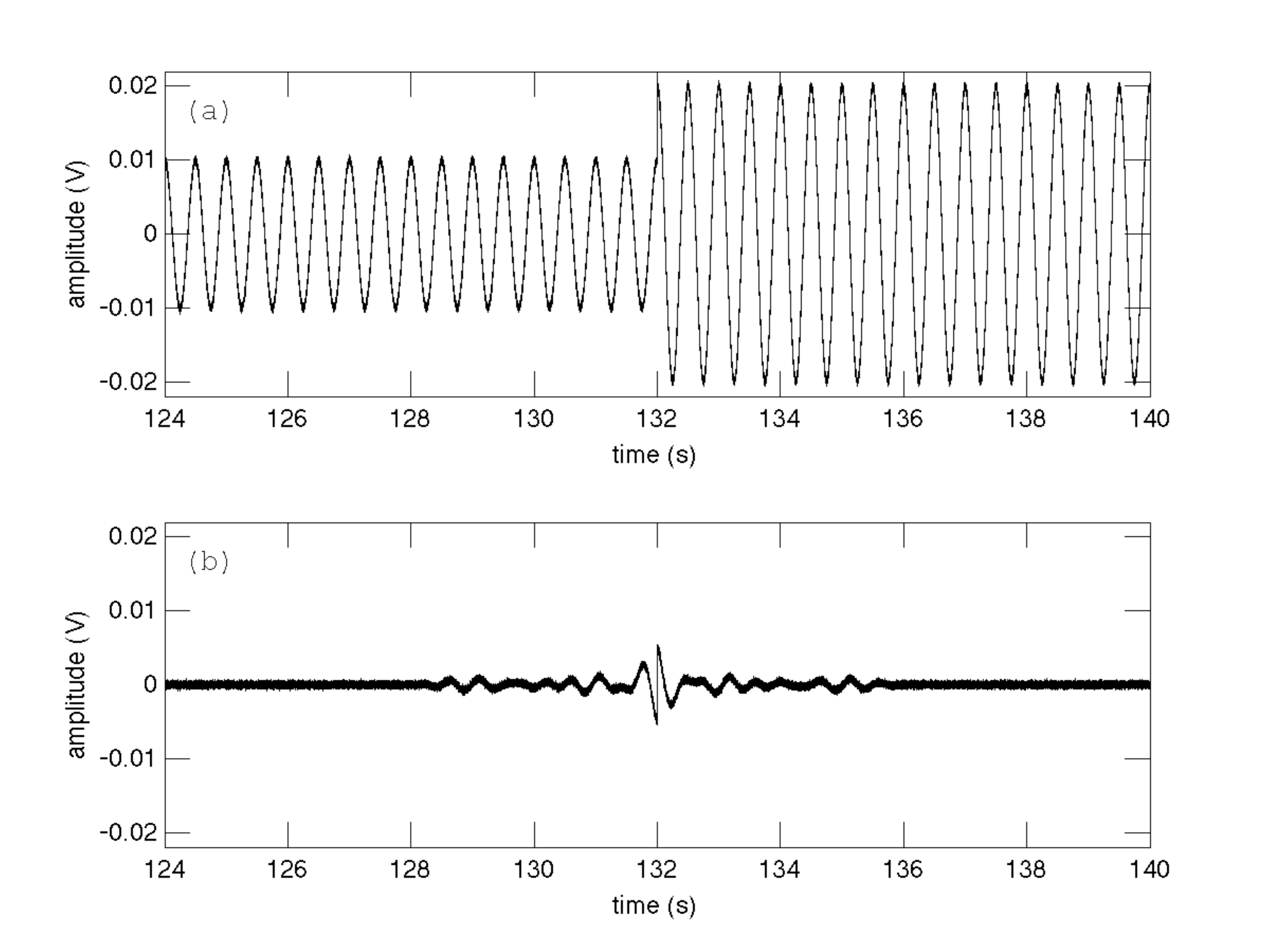}
\caption{\label{fig:stepresponse} Response of the line removal algorithm to a large
discrete step in amplitude of a line during subtraction. The line frequency
is 2\,Hz. Plot (a) shows the wave fed to the line removal algorithm, consisting of 
the oscillation having an RMS amplitude of 7\,mV before 132\,s and 14\,mV after
this time, plus pseudorandom gaussian distributed noise with an RMS of  0.2\,mV
throughout. Plot (b) shows the data after line subtraction in the 16 seconds surrounding
the discontinuity. The effects of the discontinuity are confined to the 8 second interval
around the discontinuity. Note that the amplitude of the line relative to the noise is far
greater than in the engineering data so that the effect of the discontinuity is clear
above the noise background.}
\end{center}
\end{figure}

\subsection{Response of the algorithm to injected signals}
\label{sec:signals}

Figure \ref{fig:timesig} shows the response of the subtraction algorithm to a representative
sample of the test data, but in addition a single short time duration sine Gaussian burst
has been added to the data. The ratio of the peak signal amplitude to the R.M.S. 
amplitude of the data was 3.125. Plot (a) of this figure shows the added burst. This particular burst has been selected as a worst-case scenario, since the frequency of the oscillations inside the envelope is 828~Hz, exactly equal to a dominant line frequency in the data. Plots (b), (c) and (d) show, respectively, the input data after the test signal has been added,
the correction signal generated from the data for the same time segment, and the result
of subtraction of this correction signal. The signal survives in the line subtracted data,
but the persistent fluctuations from background noise are removed. This separation is
possible due to the response time of the filter. The Fourier components are measured
over a time duration of $\tau=8$ seconds, so that any fluctuations in the line parameters on a 
timescale much shorter than $\tau$ are not significantly corrected by the filter.
This is a critical feature of this algorithm for line tracking - the response time of the line
tracker can be set based on knowledge of the properties of the noise (the stability of 
the line noise background) and the signal, the expected time duration of any transient
signals that one may wish to avoid suppressing.

Signal injections were next made at 500 random times in a 256 second data segment,
where the amplitude of the injection was varied between 0.78 and 15.23 times the
r.m.s. noise level. The mean separation between injections was approximately 0.5
seconds, an average of 16 injections per Fourier transform duration. After signal
injection, the line subtraction algorithm was applied, then the maximum height of 
the data in the domain of each injected signal was measured. This maximum height
was plotted against the height of the original injected signal in Figure \ref{fig:multiinject}.
At small injection amplitudes, the maximum height after injection reflects the noise
background, since the probability of random noise fluctuations under the signal 
leading to at least one bin that exceeds the original injection amplitude increases as
the signal to noise ratio decreases. At large injection amplitude, the ratio of amplitudes
after to before injection fluctuates about one. This behaviour is consistent with the 
maximum height bin in the line subtracted data being that having the maximum 
amplitude in the injected signal, but to this maximum height is added a noise term
that either increases or decreases the overall amplitude by an amount having 
the probability distribution of the noise background. 

The response of the line removal algorithm to a discontinuous step in the parameters
of a line was also investigated. Figure \ref{fig:stepresponse} shows the results of this
investigation. The input data, consisting of a 2\,Hz sine wave having an RMS amplitude
of 7\,mV, had its rms amplitude stepped discontinously to 14\,mV at a time of 132 seconds
on this scale. Plot (b) shows the output of the filter. Note that the input waveform has been
delayed by 4 seconds to account for the time delay between the input signal and the 
correction signal discussed in Section \ref{sec:lineremoval}, so that the time axis on both
plots represents the time at which the signal correction was applied to the data, 4 seconds
after the input data that generated the correction. The acausality of the filter
is clearly visible in plot (b). The injection time of the discontinuity was at 128 seconds, 
corresponding to a correction time of 132 seconds, but before this time and after 128 seconds
the effects of the discontinuity are already apparent in plot (b). Note, however, that the 
effects of the discontinuity are completely confined to the 8 seconds surrounding the 
correction time, because the Fourier coefficients used to evaluate the line parameters
are evaluated over $\tau=8$\,s.

\section{Summary}
\label{sec:summary}
The line subtraction method discussed in this paper successfully removes at least 76\% of
the peak amplitude of all frequency-stable lines to which it has been applied in test data. For lines for which a high signal to noise ratio can be achieved, over 99\% of the line amplitude can be removed. Tests so far have focussed on data acquired at a 
sampling rate of 16,384\,Hz, because this is the sampling frequency of fast channels in the 
GEO and LIGO data acquisition systems. At this sampling frequency, the algorithm can 
subtract 1700 separate frequencies in parallel on a single CPU, 
and still keep up with the incoming data rate.
If the rate were 16.384\,MHz, at least 1 frequency subtraction would still be possible in 
real time. Because the algorithm operates on each frequency component separately, it
is amenable to parallelization, in which case as long as at least one frequency can be 
processed in real time on each CPU, the number of frequencies subtracted depends only
on the number of CPUs.

Applications of line-removal in gravitational wave data analysis are wide-ranging. This technique
could be used as a pre-processing step for existing burst search algorithms, and may
also be useful in searches for other types of signals. Furthermore,
it presents the prospect of obtaining a very white spectrum of time domain strain data
in the gravitational wave channel. If applied successfully,
the spectral density would be frequency independent. This would allow for more robust time domain searches for unmodelled bursts, where the search sensitivity is strongly dependent only
on the amplitude of a burst signal and only weakly dependent on its shape.
The method is also applicable to other
problems where it is desirable to monitor the amplitude and phase of a wave of known
frequency, sample by sample, online. Applications in this category will be discussed in a
future publication.

\section{Acknowledgements}

Support for this work was through a Marie Curie international reintegration grant 
no. FP6-006651. The authors would also like to thank the operators of the GEO 600
instrument for furnishing the data used to test this algorithm. This paper was assigned
LIGO document number P080013.

\appendix

\section{Proof of the EFC evolution formula for Fourier coefficients}
\label{sec:appendix}

This proof starts with equation \ref{eqn:fourier} for the $k^{\mathrm{th}}$
Fourier coefficient, where we separate the $j=0$ term from the sum
and write it at the end.

\begin{equation}
\mathcal{F}_k(x_0,~x_1,\cdots,x_{N-1})=
\sum_{j=1}^{N-1}x_j e^{\frac{-2\pi i jk}{N}} + x_0.
\label{eqn:separate}
\end{equation}

\noindent
Next write down an expression below for
the $k^{\mathrm{th}}$ Fourier coefficient of the data forward timeshifted
by one sample, where the new sample is $x_N$. Rearrange this expression
by separating the term in $x_N$ and factor out a common phase.

\begin{eqnarray}
\mathcal{F}_k(x_1,~x_2,\cdots,x_N)&=&
\sum_{j=1}^N x_j e^{\frac{-2\pi i(j-1)k}{N}} \\ \nonumber
&=&\sum_{j=1}^{N-1}x_j e^{\frac{-2\pi i(j-1)k}{N}}
+ x_N e^{\frac{-2\pi i (N-1) k}{N}}.\\ \nonumber
&=&e^{\frac{+2\pi ik}{N}}\left(
\sum_{j=1}^{N-1}x_j e^{\frac{-2\pi ijk}{N}}
+ x_N e^{\frac{-2\pi iNk}{N}}
\right) \\ \nonumber
&=&e^{\frac{+2\pi ik}{N}}\left(
\sum_{j=1}^{N-1}x_j e^{\frac{-2\pi ijk}{N}} + x_N
\right).\\ \nonumber
\label{eqn:evolve}
\end{eqnarray}

\noindent
The expression in parentheses can be related to $\mathcal{F}_k(x_0,~x_1,\cdots,x_{N-1})$
by using equation \ref{eqn:separate}.

\begin{equation}
\mathcal{F}_k(x_1,~x_2,\cdots,x_N)=
e^{\frac{+2\pi ik}{N}}\left(
\mathcal{F}_k(x_0,~x_1,\cdots,x_{N-1})+(x_N-x_0)\right).
\label{eqn:evolution}
\end{equation}

\noindent
This is the desired result.

\end{document}